\documentclass{article}

\usepackage{arxiv}

\usepackage[utf8]{inputenc} 
\usepackage[T1]{fontenc}    
\usepackage{hyperref}       
\usepackage{url}            
\usepackage{booktabs}       
\usepackage{amsfonts}       
\usepackage{nicefrac}       
\usepackage{microtype}      
\usepackage{amsmath}
\usepackage{graphicx}
\usepackage{caption}
\usepackage{float}

\DeclareCaptionFormat{myformat}{#1#2#3\hrulefill}
\title{Motion artifact removal and signal enhancement to achieve in vivo dynamic full field OCT}

\author{
  Jules Scholler\thanks{\url{https://www.jscholler.com}} \\
  Institut Langevin \\
  ESPCI Paris, CNRS, PSL University \\
  1 rue Jussieu, 75005 Paris, France\\
  \texttt{jules.scholler@espci.fr} \\
}

\begin{document}
\maketitle

\begin{abstract}
We present a filtering procedure based on singular value decomposition to remove artifacts arising from sample motion during dynamic full field OCT acquisitions. The presented method succeeded in removing artifacts created by environmental noise from data acquired in a clinical setting, including in vivo data. Moreover, we report on a new method based on using the cumulative sum to compute dynamic images from raw signals, leading to a higher signal to noise ratio, and thus enabling dynamic imaging deeper in tissues.
\end{abstract}

\section{Introduction}

Optical coherence tomography (OCT) is routinely used for 3D imaging of microstructures in tissue and relies on the endogenous backscattering contrast \cite{Huang_91, drexler_optical_2015}. Full-field optical coherence tomography (FFOCT) is an \textit{en face}, high transverse resolution version of OCT \cite{Beaurepaire_98, Dubois_04}. Using a camera and an incoherent light source, FFOCT acquires \textit{en face} virtual sections of the sample at a given depth and has been used for biology \cite{Benarous} and medicine \cite{Jain}. Recently, a novel contrast mechanism has been exploited by measuring temporal fluctuations of the backscattered light in a technique called dynamic full field OCT (D-FFOCT) \cite{Apelian_16}. In ex vivo fresh tissues, these dynamic measurements reveal subcellular structures that are very weak backscatterers and provide contrast based on their intracellular motility \cite{Leroux_16, Thouvenin_motility}. A similar technique is used in regular OCT for retinal angiography called OCTA where speckle variance is analyzed on several B-Scans (typically 8 frames) to produce binary images of the retinal vasculature \cite{Kashani_17}. Due to the high spatial resolution ($<1~\mu m$) and the number of frames used in D-FFOCT (typically 512), our method is not only sensitive to capillaries but also to intracellular motility signals and produces a contrast that reveals living cells \cite{Scholler_19}. The penetration depth of D-FFOCT is typically ten times less than FFOCT due to the  small cross-section of the moving scatterers leading to weak signals, limiting its use in thick samples. Up to now, using D-FFOCT for in vivo imaging has remained elusive as this technique is sensitive to nanometric axial displacements of the sample. The same problems arise for OCTA and several approaches have been developed to remove bulk motion of the eye \cite{Jia_12,deCarlo2015}. In this paper we propose two methods to overcome the aforementioned limitations.

First, we introduce a framework based on the singular value decomposition (SVD) to filter out the axial displacement of the sample from the local fluctuations linked to intracellular motility, enabling in vivo use of D-FFOCT. SVD based algorithms have been previously applied to OCT data, e.g. for smart OCT where the SVD is applied to the reflection matrix in order to extend the penetration depth \cite{Badon_svd_16}. An SVD filtering method for D-FFOCT has been previously proposed for simulated data \cite{Ammari_svd} but does not work on our experimental data mainly because the image formation model is different. Here we propose to find eigenvectors associated with axial motion and filter them out. Similar SVD based algorithms for spatio-temporal filtering have been used effectively in acoustics for Doppler acquisitions \cite{SVD_doppler,SVD_doppler2}. In each case the goal is to use the SVD to transform the initial data in a new basis that is more suitable for filtering and identifying outliers. As opposed to \cite{SVD_doppler, SVD_doppler2}, our approach reconstructs the signals in the initial space before computing the dynamic image rather than constructing the image in the SVD space. The main advantage of using the SVD rather than Fourier analysis here is that the filter adapts to the data set which exhibits different amounts of artifacts with random patterns.

Secondly, we present a new operator to compute the local dynamics based on the cumulative sum in order to enhance the non-stationary part of the signal, leading to a great increase in the signal to noise ratio (SNR).

Finally, we report on the first D-FFOCT acquisition in vivo to image the mouse liver at $80~ \mu m$ depth where the two proposed algorithms greatly improved the image quality by removing motion artifacts and increasing the SNR by a factor of 3.

\section{Removing artifacts using SVD}

\begin{figure}
    \centering
    \includegraphics[width=0.7\linewidth]{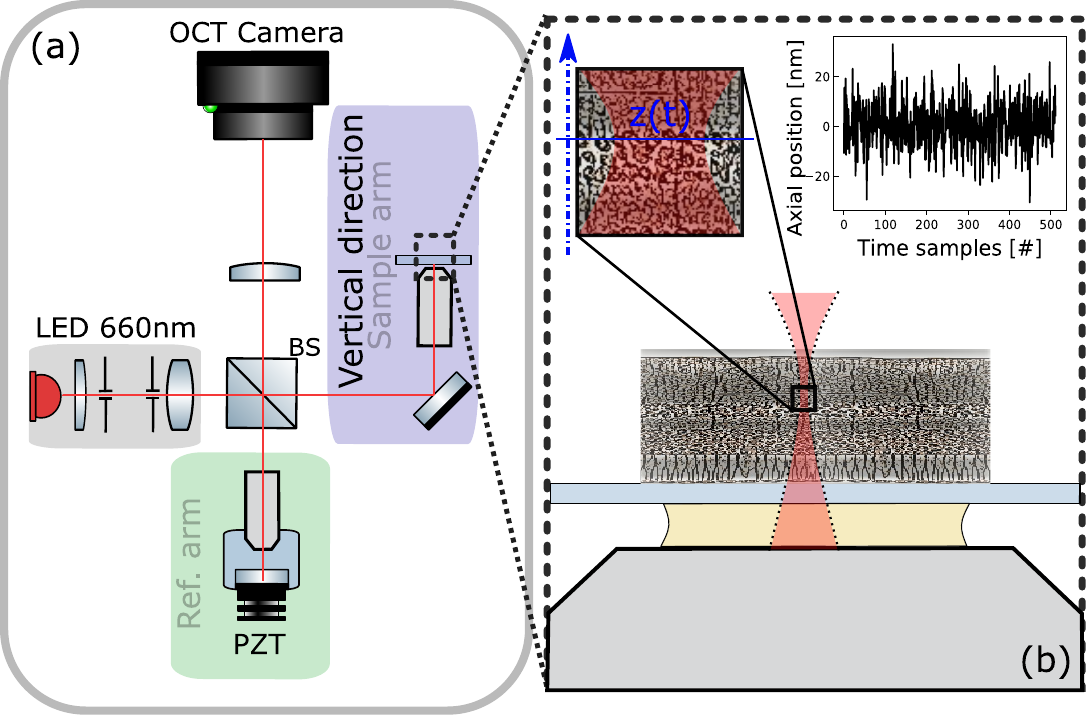}
    \caption{(a) FFOCT setup in inverted configuration \textit{top view}. Microscope objectives are Olympus UPlanSApo 30x 1.05 NA. OCT Camera: ADIMEC Q-2A750-CXP. Light source: Thorlabs M660L3. - PZT: piezoelectric translation - BS: 50/50 beam splitter. (b) Schematic of the sample axial oscillations around the coherence volume due to mechanical vibrations, simulated on a graph in the top right corner. The setup is illustrated with oil immersed objectives where the probed volume depth is $1~\mu m$.}
    \label{fig1}
\end{figure}

In order to construct a D-FFOCT image, a stack of typically 512 direct images ($1440 \times 1440$ pixels) is acquired with a standard FFOCT using our custom software \cite{FFOCT_JS}. The FFOCT setup consists of a Linnik interferometer where both arms contain identical microscope objectives. The reference arm contains a silicon mirror mounted on a piezoelectric translation (PZT) used for phase modulation. In a typical FFOCT experiment, at least two images are acquired with different phase modulations and the FFOCT image is constructed by using appropriate phase-shifting algorithms \cite{Dubois_04,Scholler_19}. For D-FFOCT experiments the PZT position is not modulated, fluctuations arise by scatterers motions inside the coherence volume. In this paper we used data acquired from two different setups. The first one is a laboratory setup shown in Fig.~\ref{fig1} and the second one is a LightCT commercial setup manufactured by LLTech SAS. The characteristics of both of these setups are summarized in Table~\ref{table1}. In the first report on D-FFOCT, the level of the dynamic signal at each pixel was computed using a running standard deviation averaged over the whole acquisition \cite{Apelian_16} so that each pixel is processed independently. Calculating the standard deviation of the signal in time removes highly scattering stationary structures such as collagen or myelin fibers and reveals cells with a much better contrast. Indeed, strongly backscattering structures can dominate the signal even outside the coherence volume thereby masking weakly scattering structures such as cells. In the laboratory, we succeeded in stabilizing the setup by mounting it on a sturdy optical bench, and carried out ex vivo experiments without motion artifacts. For real life applications however, D-FFOCT devices are currently being used by clinicians in hospitals for imaging biopsied tissues \cite{LLTech_spie_19} and by anatomo-pathologists in busy environments with vibrations arising from vibrational modes of the building, from people walking around the device and from air conditioning. Mechanical vibrations can lead to sample arm motion or oscillations, creating strong signal fluctuations, especially from highly reflective structures such as collagen fibers.

\begin{table}
\caption{\textbf{Setup Characteristics.}}
\label{table1}
\centering
\begin{tabular}{c|c|c|c}
Setup & Lab. setup Fig.~\ref{fig1}(a) & LightCT (LLTech SAS) \\ \hline
Transverse resolution $[\mu m]$   &  0.4   &    1.5             \\
 Axial resolution $[\mu m]$ &       1     &     1              \\ 
 Field of view $[\mu m \times \mu m]$ & $390\times 390$ & $1260\times 1260$ \\
 Framerate $[Hz]$ & 150 & 150 \\
\end{tabular}
\end{table}

\subsection{Motion artifact model}

The intensity recorded by the camera is the sum of the backscattered light from both the sample and the reference arm \cite{Scholler_19}:
\begin{align}
    I(\boldsymbol{r},t) = \eta \frac{I_0}{4} \left(  R(\boldsymbol{r},t) + R_{inc} + R_{ref} + 2\sqrt{R(\boldsymbol{r},t) R_{ref}}cos \left( \Delta \phi (\boldsymbol{r},t)) \right) \right)
\end{align}
where $I(\boldsymbol{r},t)$ is the intensity recorded at position $\boldsymbol{r}=(x,y)$ and time $t$, $\eta$ is the camera quantum efficiency, $I_0$ is the power LED output impinging on the interferometer considering a 50/50 beam-splitter, $R_{ref}$ is the reference mirror reflectivity (i.e. the power reflection coefficient), $R(\boldsymbol{r},t)$ is the sample reflectivity (i.e. the power reflection coefficient) at position $\boldsymbol{r}$ and time $t$, $\Delta \phi(\boldsymbol{r},t)$ is the phase difference between the reference and sample back-scattered signals at position $\boldsymbol{r}$ and time $t$, $I_{incoh} = R_{inc}I_0/4$ is the incoherent light back-scattered by the sample on the camera, mainly due to multiple scattering and reflections out of the coherence volume. The dynamic signal is computed as the average of the running temporal standard deviation and the processed dynamic signal can be written:
\begin{align}
    I_{dyn}(\boldsymbol{r}) =\frac{1}{N} \sum_i SD\left(\frac{\eta I_0}{2}\sqrt{R_s(\boldsymbol{r},t_{[i,i+\tau]}) R_{ref}}cos\left(\Delta \phi_s(\boldsymbol{r},t_{[i,i+\tau]})\right)\right)
\end{align}
where $SD$ is the standard deviation operator, $N$ is the total number of sub-windows, $\tau$ is the sub-windows length so that $t_{[i,i+\tau]}$ is the time corresponding to one sub-window, $R_s$ and $\Delta \phi_s$ are respectively the reflectivity and phase of the local scatterers that induce the temporal fluctuations that D-FFOCT aims to measure. In the event of small displacements of the entire sample, on the order of the depth of field or smaller, the processed signals will be the sum of the actual local fluctuations and the modulation created by the bulk sample motion creating a global phase shift. The resulting artifacts are therefore proportional to the sample reflectivity, which is orders of magnitude higher than the reflectivity of the scatterers probed by D-FFOCT (e.g. mitochondria and vesicles) leading to strong artifacts on the dynamic image, which mask the signal of interest. In the presence of mechanical noise, the measured fluctuation can be written:
\begin{align}
    I_{mes}(\boldsymbol{r}) = I_{dyn}(\boldsymbol{r}) + I_{art}(\boldsymbol{r})
\end{align}
where the artifactual signal can be expressed as:
\begin{align}
    I_{art}(\boldsymbol{r}) = \frac{1}{N} \sum_i SD\left(\frac{\eta I_0}{2}\sqrt{R_s(\boldsymbol{r},t_{[i,i+\tau]}) R_{ref}}cos\left(\frac{4\pi}{\lambda}z(t_{[i,i+\tau]})\right)\right)
\end{align}
where $z(t_{[i,i+\tau]})$ is the sample axial displacement on the $i-th$ sub-window. Here we neglected the sample deformation for the sake of clarity. Nonetheless it could be taken into account by processing the stack in spatial patches where deformations are negligible. For a highly reflective zone we have $I_{dyn}(\boldsymbol{r}) \ll I_{art}(\boldsymbol{r})$ and the dynamic signal is completely masked by artifacts that look like the corresponding static FFOCT image that could be obtained by randomly sampling the path difference instead of using the standard PZT modulation mentioned before. Indeed, modulating the position of the sample around the coherence volume is equivalent to modulating the piezo position, which explains why artifacts look like the standard FFOCT image.

\subsection{Proposed algorithm}

\begin{figure}
    \centering
    \includegraphics[width=0.8\linewidth]{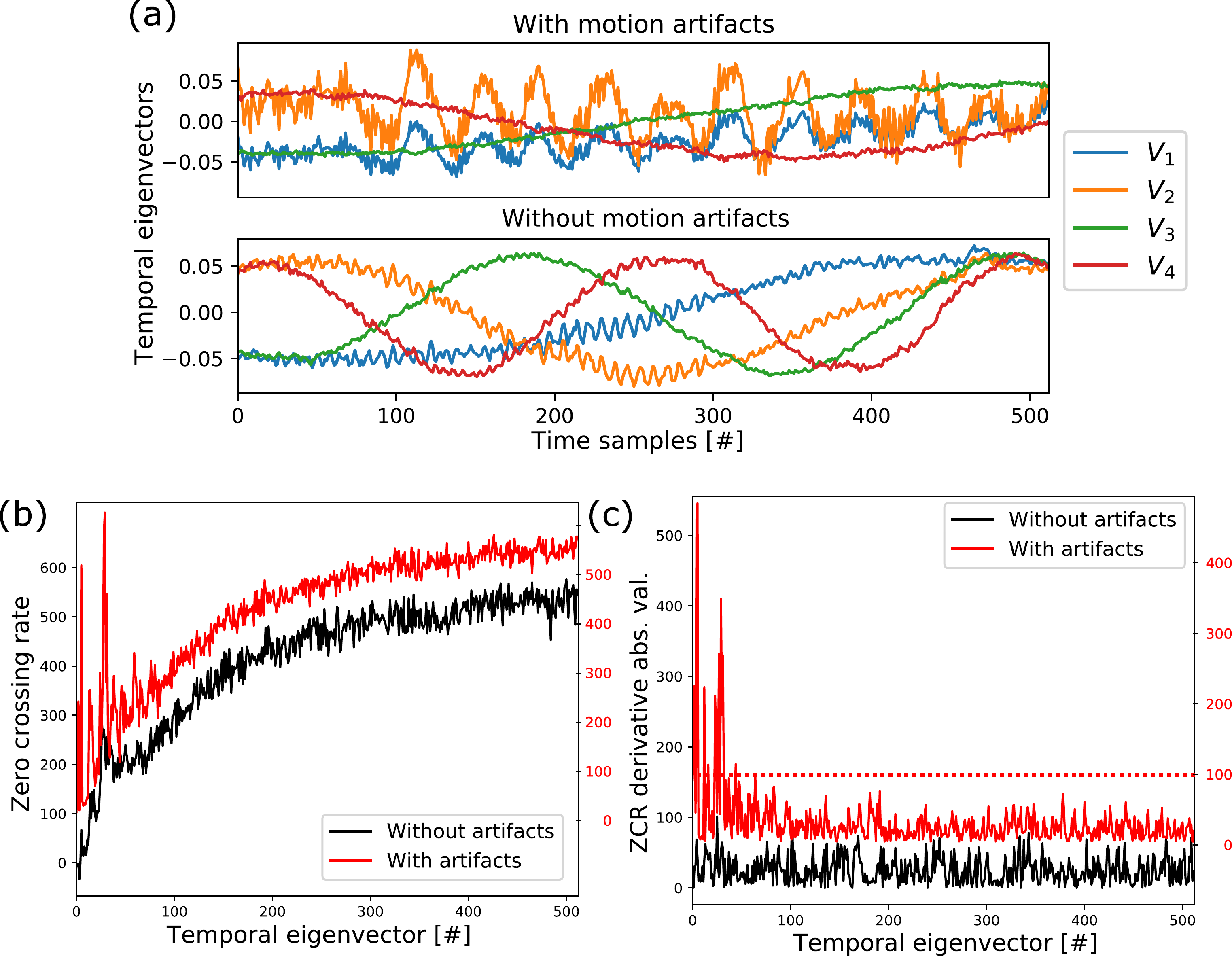}
        \caption{(a) First few temporal eigenvectors. \textit{Top:} for an acquisition with motion artifacts where $V_1$ and $V_2$ were detected as motion artifact and removed. \textit{Bottom:} for an acquisition without visible artifacts. (b) Zero crossing rate computed for each temporal eigenvectors. (c) Absolute value of the derivative of the zero crossing rate computed for each temporal eigenvector. Artifacts were detected by thresholding the curve above 3 standard deviations. The baseline was arbitrarily increased for the red curves in (b) and (c) in order to increase readability.}
    \label{fig2}
\end{figure}

In order to remove motion artifacts we want to use the SVD as an adaptive filter that could separate motility signals from motion induced signals. The first step is to unfold the 3D $M(x,y,t)$ cube of data into a 2D matrix $M_u(\boldsymbol{r},t)$ to perform the decomposition. Higher dimensions of SVD do exist but are not required here as the horizontal $x$ and vertical $y$ dimension do not differ when considering axial motion artifacts. SVD is the generalization of the eigendecomposition of a positive semidefinite normal matrix, and can be thought of as decomposing a matrix into a weighted, ordered sum of separable matrices which will become handy when reconstructing the SVD-denoised signals:
\begin{align}
    M_u = U\Sigma V^\star = \sum _ { i } \sigma _ { i } \mathbf { U } _ { i } \otimes \mathbf { V } _ { i }
\end{align}
where $\otimes$ is the outer product operator, $U$ contains the spatial eigenvectors, $V$ contains the temporal eigenvectors and $\Sigma$ contains the eigenvalues associated with spatial and temporal eigenvectors. Performing the SVD on an unfolded $M_u(1440\times 1440,512)$ dynamic stack takes around 30 seconds on a workstation computer (Intel i7-7820X CPU, 128 Gbyte of DDR4 2666 MHz RAM) and requires 45 GByte of available RAM using LAPACK routine for SVD computation without computing full matrices. Investigating such decomposition for artifact-free data sets we found that spatial eigenvectors related to motion artifacts have particular and easily identifiable associated temporal eigenvectors. Indeed, when looking at the first temporal eigenvectors, we observed sinusoid-like patterns with increasing frequency, see Fig.~\ref{fig2}(a). In presence of motion artifacts, temporal eigenvectors appeared with random, high-frequency components that are easy to detect with simple features. Here, the zero crossing rate (the number of times a function crosses $y=0$) is used to detect temporal eigenvectors involved in motion artifacts. In presence of motion artifacts some of the firsts temporal eigenvectors present a high zero crossing rate, see Fig.~\ref{fig2}(b) and \ref{fig2}(c). In order to detect these outliers we computed the absolute value of the derivative of the zero crossing rate (D-ZRC $= |ZRC_{i+1}-ZRC_{i}|$) and applied a threshold: if the D-ZRC is higher than three times the D-ZRC standard deviation then the corresponding eigenvalue $\sigma_i$ is set to zero in $\hat{\Sigma}$ and the SVD-denoised stack $\hat{M}_u$ is reconstructed as:
\begin{align}
    \hat{M}_u = \sum _ { i } \hat{\sigma _ { i }} \mathbf { U } _ { i } \otimes \mathbf { V } _ { i }
\end{align}
The SVD-denoised stack $\hat{M}_u$ can then be folded back to its original 3D shape $\hat{M}$ and the dynamic computation can be performed. Interestingly, the use of an automatic selection of eigenvectors allows a more reproducible analysis. For example, the SVD can be performed on spatial sub-regions without visible artifacts, something very hard to obtain with manual selection of eigenvectors. This can also improve the filtering procedure in the case of sample spatial deformation or if the computation requires too much RAM. 

\subsection{Results}

\begin{figure}
    \centering
    \includegraphics[width=\linewidth]{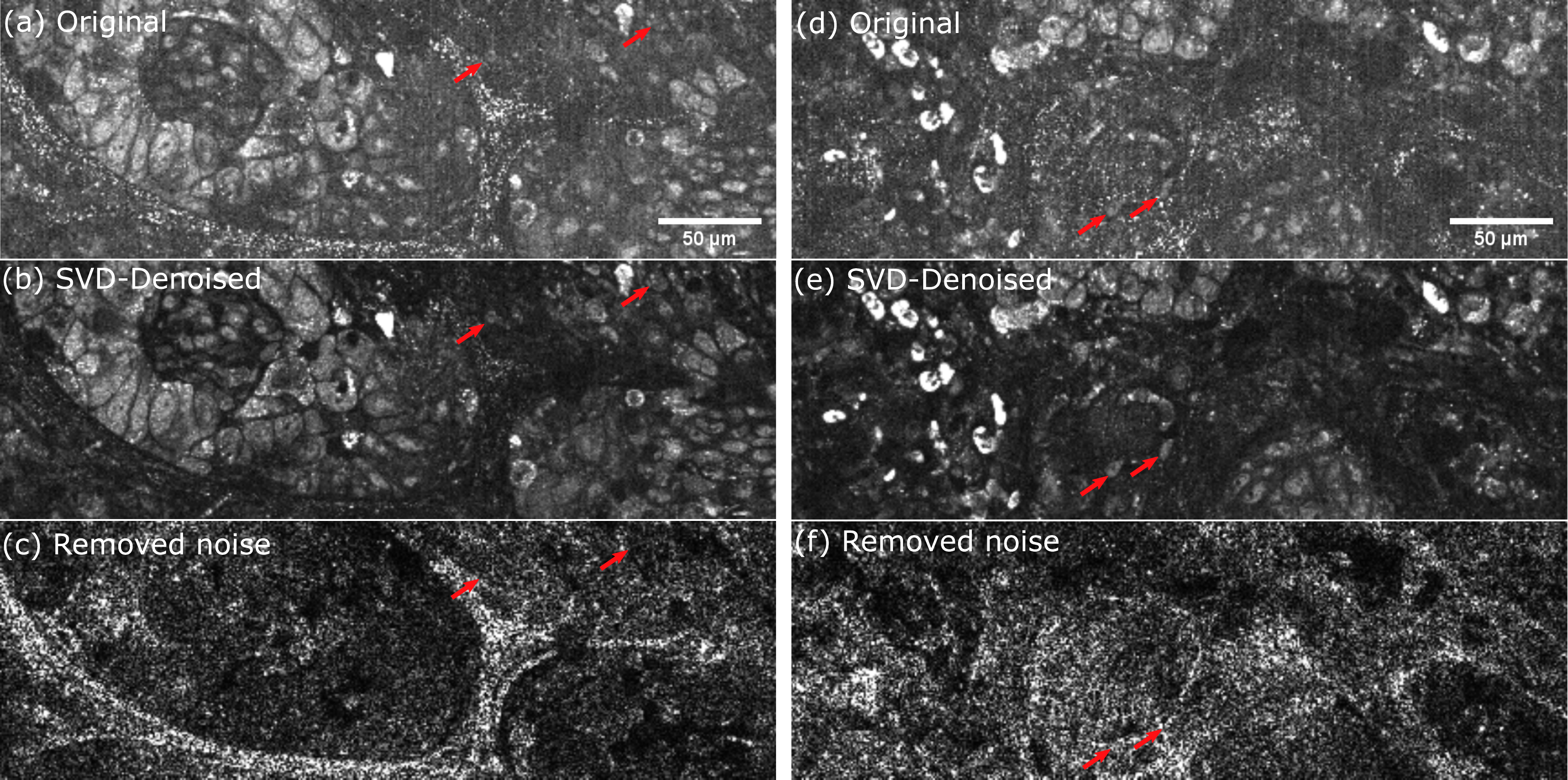}
    \caption{Lung biopsy for cancer detection taken on the LLTech clinical setup. Artifacts arise mainly from mechanical vibration and air conditioning. (a)(d) Original D-FFOCT images computed on the raw stack. (b)(e) Denoised images computed with SVD filtering. (c)(f) Sum of the spatial eigenvectors absolute value removed by the SVD filtering. Red arrows are highlighting cells that were partially masked by motion artifacts.}
    \label{fig3}
\end{figure}

\begin{figure}
    \centering
    \includegraphics[width=0.45\linewidth]{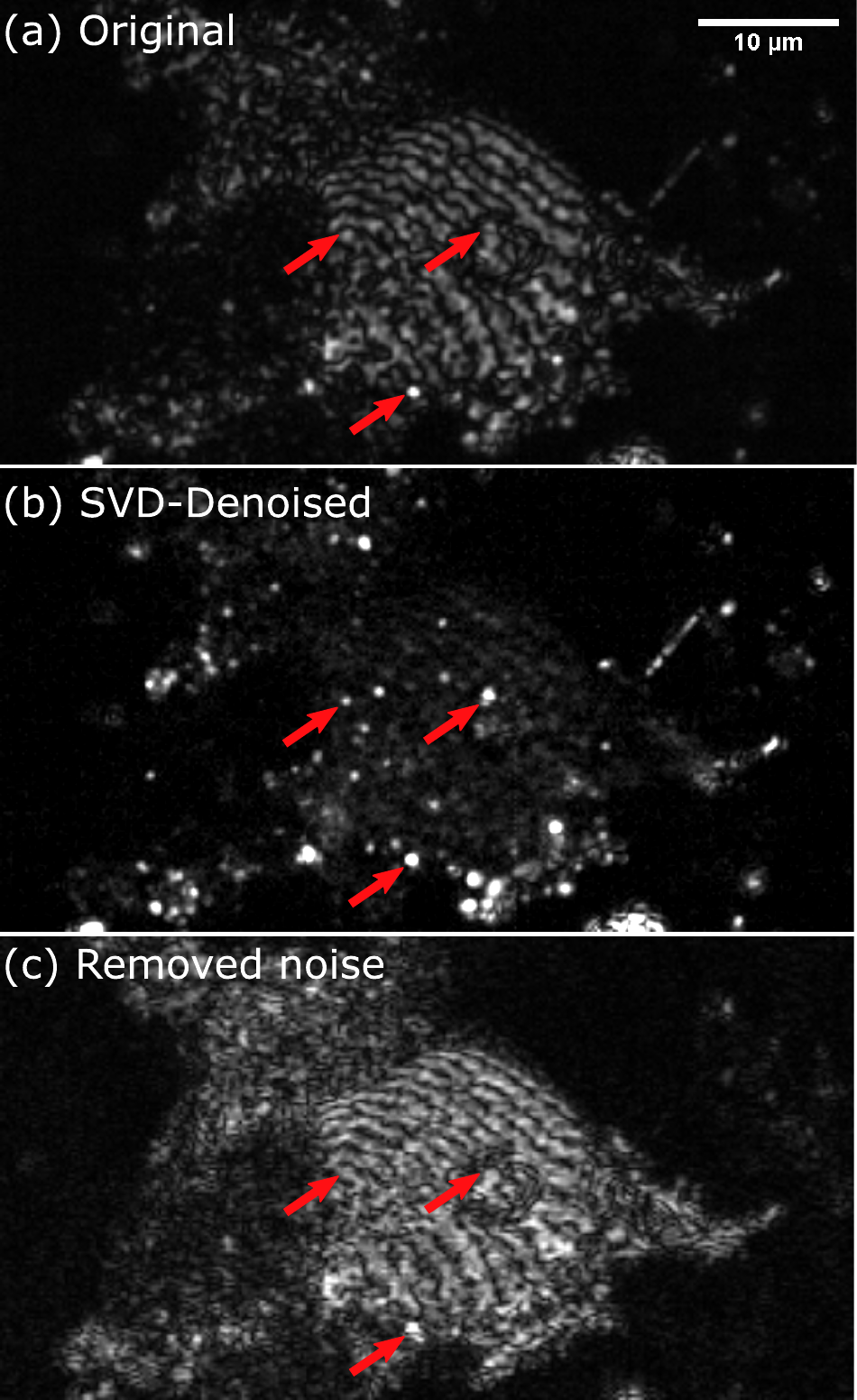}
    \caption{Fibroblast image taken on the setup presented Fig.~\ref{fig1}. Artifacts arise mainly from mechanical vibrations leading to fringes pattern. (a) Original D-FFOCT image computed on the raw stack. (b) SVD-denoised image computed with SVD filtering. Subcellular features appear with a much better contrast enabling segmentation and tracking. (c) Sum of the spatial eigenvectors absolute value removed by the SVD filtering. Red arrows are highlighting subcellular features, only the bottom one was visible on the original image.}
    \label{fig4}
\end{figure}

We tested the proposed SVD filtering on different acquisitions taken on different setups (the one presented in Fig.~\ref{fig1}(a) and the LightCT system manufactured by LLTech SAS). When motion artifacts were present, image quality after denoising was greatly improved in each case. In Fig.~\ref{fig3} we present lung biopsy images taken with the LightCT system in a clinic, where imaged tissues were waste tissues from biopsy procedures that were destined to be destroyed, and we imaged them just before destruction. The imaging was carried out according to the tenets of the Declaration of Helsinki and followed international ethic requirements for human tissues. SVD filtering effectively removes motion artifacts from collagen fibers and reveals cells in Fig.~\ref{fig3}(b) and \ref{fig3}(e). D-FFOCT images were also higher contrast after SVD denoising, allowing easier interpretation for clinical applications, e.g. lung tumor detection in the presented images. We imaged fibroblasts with the setup presented in Fig.~\ref{fig1}(a). Cells were very flat leading to fringes pattern created by the specular reflection on their surface. These fringes were highly visible on the processed D-FFOCT image preventing the visualization of subcellular structures, see Fig.~\ref{fig4}(a). After SVD filtering it is possible to distinguish single subcellular entities and track them, see Fig.~\ref{fig4}(b), enabling biological study without the need of a costly optical bench setup.

\section{Extending penetration depth using non-stationarities}

In addition to motion artifacts, a drawback of D-FFOCT compared to standard static FFOCT is the reduced penetration depth. While FFOCT can acquire images as deep as $1~mm$, D-FFOCT is limited to about $100~\mu m$ due to the weak signal level produced by the sample fluctuations we wish to measure. In order to enhance the dynamic signal strength and so improve penetration depth, we propose computation of the dynamic image from the cumulative sum of the signal, rather than the raw signal. Indeed, the model for the dynamic image formation is small scatterers moving in the coherence volume during the acquisition leading to phase and amplitude fluctuations in the conjugated camera pixel. While a pure Brownian motion is stationary, hyper-diffusive displacements are not and we therefore propose to use the cumulative sum to enhance these non-stationarities. Intuitively, summing a centered noise will give a noisy trajectory that stays close to zero whereas if there is a small bias it will be summed for every sample and the cumulative sum will therefore have a slope equal to this bias.

\subsection{Theoretical considerations}

Let us consider an array of random values drawn from a zero centered Gaussian distribution. If the number of samples is large the mean of the array will be close to zero and equivalently the sum of all the samples will also be close to zero. Taking the cumulative sum of such an array gives a so called \textit{Brownian bridge} (the curve starts and ends close to zero and makes a "bridge" between these two points). Theoretically, Brownian bridges are expected to be maximal close to the edges as the probability distribution of the maximum is the third arcsin law which has a typical U-shape  \cite{Levy_brownian_bridge}. More importantly, the Brownian bridge maximum follows a Rayleigh distribution. If we consider a Brownian bridge $W_s ~ \forall ~ s \in [0,1]$:
\begin{align}
    W_M=sup\{W_s:s\in [0,1]\} \\
    \mathbb { P } [ W_M \leq u ] = 1 - e^{ - 2 u ^ { 2 }}
\end{align}
where $W_M$ is the supremum of the bridge and $\mathbb { P } [ W_M \leq u ]$ is the probability of the supremum being less than $u$. According to the Rayleigh distribution, the Brownian bridge maximum must therefore scale as $\sqrt{t}$ with $t$ scaling as the number of frames. Now, if there is a bias in the distribution, which is the case if a scatterer is moving with constant velocity in the coherence volume, the cumulative sum will scale as $\frac{t}{2}$ due to the slope introduced by the bias. It will also be either always positive or negative and the maximum will be reached around the center of the bridge. The cumulative sum will therefore exhibit a completely different behavior for centered noise than for actual motility signals, leading to a better signal to noise ratio on dynamic images. Note that for Brownian bridges it is often observed that the function changes sign regularly (the probability of sign changes is also well established \cite{Levy_brownian_bridge}), which is not the case when there are non-stationarities.

We simulated an experiment by introducing a linear bias of $\sigma/3$ on a centered Gaussian distribution which is not perceptible on the signals presented in Fig.~\ref{fig5}(a). Looking at the cumulative sum the bias is much more obvious as the maximum reached by the bridge is three times higher, hence motility signals are detected with a higher sensitivity using the cumulative sum.

\subsection{Results}

\begin{figure}[H]
    \centering
    \includegraphics[trim={1cm 0cm 0cm 0cm},clip,width=\linewidth]{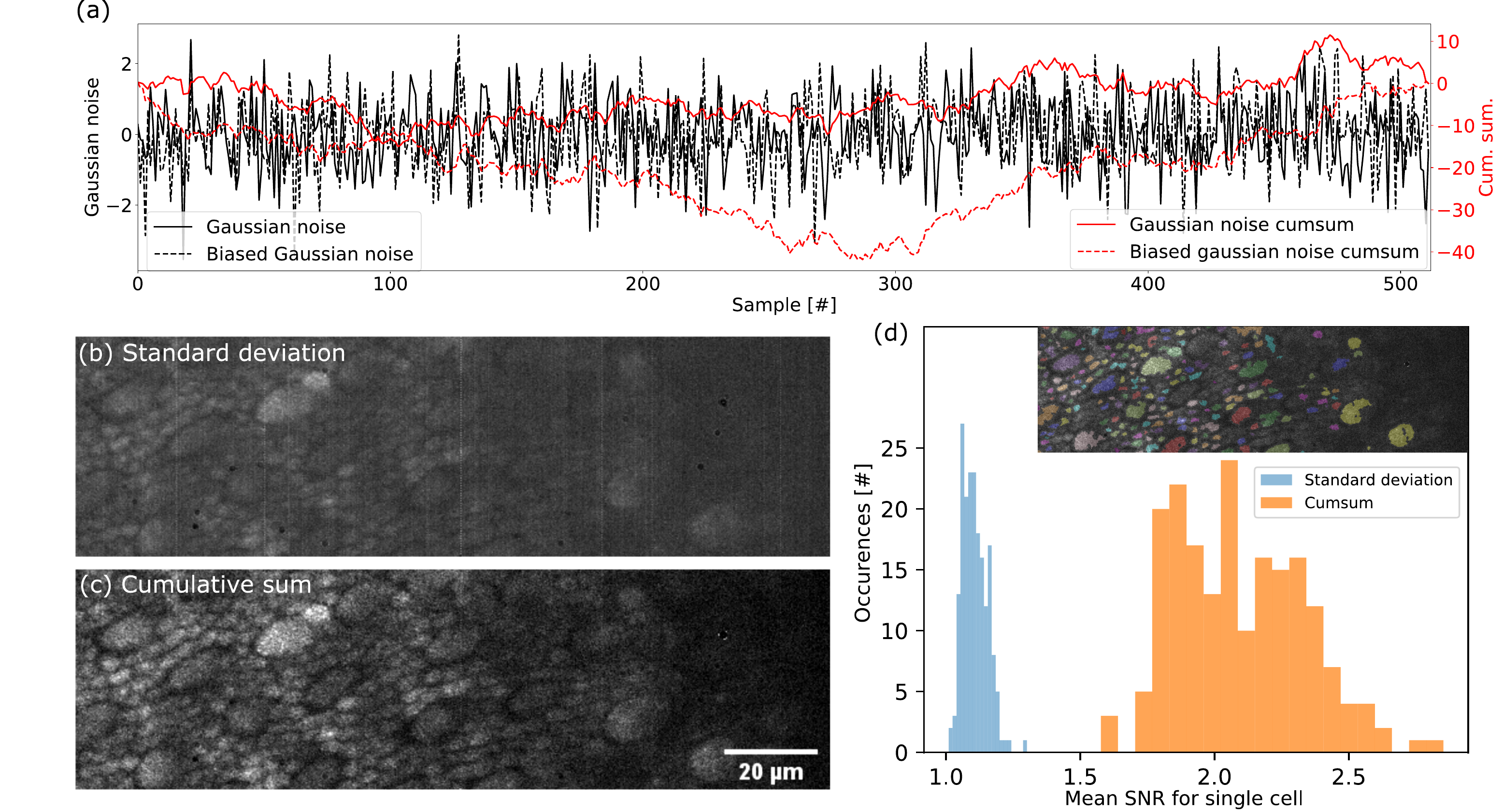}
    \caption{(a) Simulation of Gaussian noise with and without bias and their cumulative sum. The maximum reached by the cumulative sum is $3$ times higher with the bias. (b) Dynamic image of a macaque photoreceptor layer using standard deviation. (c) Dynamic image of a macaque photoreceptor layer using the proposed method based on cumulative sum. (d) Histogram of SNR for 192 single cells segmented in (b) and (c). Segmentation results are shown in the top right corner. The gain in SNR for this data set is $1.96$ on average for each cell.}
    \label{fig5}
\end{figure}

The dynamic image is computed by taking the average of the maxima of the absolute values of the running cumulative sum:
\begin{align}
    I'_{dyn}(\boldsymbol{r}) =\frac{1}{N} \sum_i max\left(|CumSum\left(I(\boldsymbol{r},t_{[i, i+\tau]})-\bar{I}(\boldsymbol{r},t_{[i, i+\tau]})\right)|\right)
\end{align}
where $CumSum$ is the cumulative sum operator, $N$ is the total number of sub-windows, $\tau$ is the sub-windows length so that $t_{[i,i+\tau]}$ is the time corresponding to one sub-window and $\bar{I}(\boldsymbol{r},t_{[i, i+\tau]})$ is the signal mean on the sub-window. We tested the proposed method with $\tau=50$ on the photoreceptor layer of an explanted macaque retina at $85~\mu m$ depth that presents a horizontal gradient of SNR, see Fig.~\ref{fig5}(b) and \ref{fig5}(c). In order to quantify the gain in SNR we segmented 192 single cells using Trainable Weka \cite{Weka_17} and computed the SNR for each cell (the SNR was computed as the mean intensity of the pixels inside the cells divided by the mean intensity of the background), see Fig.~\ref{fig5}(d). In this case the SNR is doubled with the proposed method and the camera column noise is almost completely removed. We tested the proposed algorithm on several acquisitions on tissues and cell cultures and the average SNR improvement factor was 1.9, allowing imaging deeper into tissues with D-FFOCT.

\section{Applying both methods for in vivo dynamic imaging}

\begin{figure}
    \centering
    \includegraphics[width=0.65\linewidth]{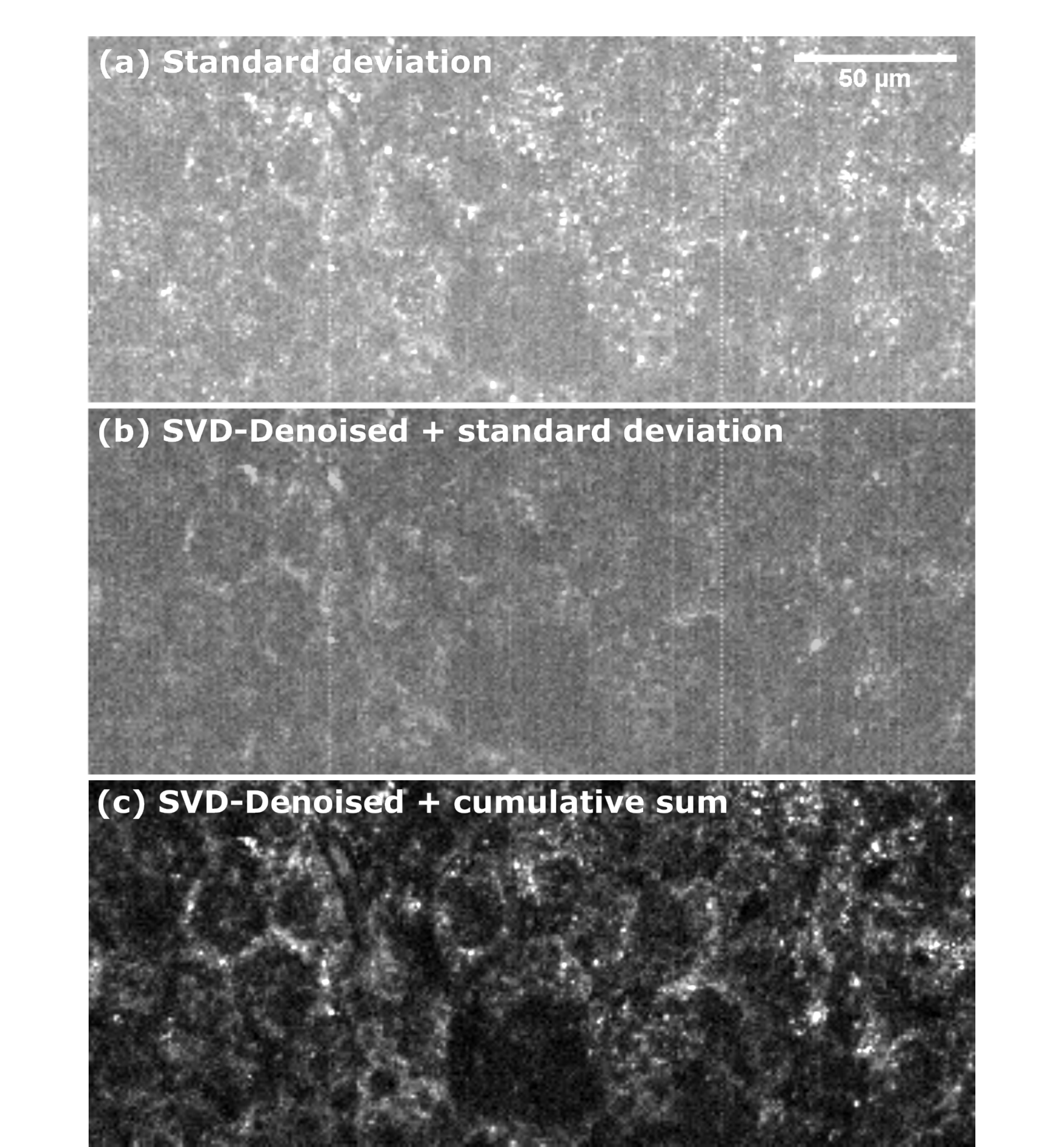}
    \caption{In vivo mouse liver dynamic image taken on LLTech clinical setup with custom mount. Artifacts mainly arise from the breathing and heartbeat. (a) In vivo D-FFOCT image of a mouse liver computed on the raw stack with the standard deviation. (b) In vivo D-FFOCT image of a mouse liver computed on the SVD-denoised stack with the standard deviation. (c) in vivo D-FFOCT image of a mouse liver computed on the SVD-denoised stack with the cumulative sum.}
    \label{fig6}
\end{figure}

The proposed SVD filtering procedure is of great interest for applying D-FFOCT in vivo for removing sample motion such as eye motion for retinal imaging. In order to limit lateral drifts and to maintain contact, a custom head was adapted on the sample arm of a LightCT setup combined with a pump to create a weak suction force. We acquired a stack of images on a living mouse liver at $80 ~\mu m$ depth. The animal manipulation protocol was approved by our local animal care committee. The mouse (4 week-old C57BL/6 (Janvier Lab, Le Genest Saint Isle, France)), was anesthetized by isoflurane, and sacrificed after the imaging procedure by $CO_2$ inhalation. The standard D-FFOCT images were very noisy mainly due to the heartbeat and breathing of the mouse, leading to tissue motion that creates artifacts, see Fig.~\ref{fig6}(a). On applying the proposed SVD filtering, we were able to remove motion artifacts Fig.~\ref{fig6}(b). Nonetheless signals are still very low due to the deep imaging in a strongly scattering organ and applying the cumulative sum algorithm dramatically increased the SNR by a factor of $3$ Fig.~\ref{fig6}(c). In the end, there are remaining artifacts produced by the coherence volume axial drift during the acquisition. Indeed, if the coherence volume shifts more than its axial extension, even if motion artifacts are perfectly removed, the probed dynamics would be averaged over several depths leading to an axial blur. To overcome this issue, the position of the coherence gate inside the sample may be compensated by monitoring the breathing and moving the reference arm with a precision corresponding to the optical sectioning ($1~\mu m$ for the in vivo acquisition here) in order to compensate for the axial drift.

\section{Conclusion}

We proposed a filtering algorithm based on the SVD to effectively remove motion artifacts from dynamic images. The proposed method adds $\sim 40$ seconds for a $(1440,1440,512)$ stack which will require GPU processing in order to speed up the process for real time applications. This method was applied on an in vivo data set and is promising as long as axial motion is smaller than the coherence volume depth. Tracking and compensating methods are currently being investigated in order to acquire D-FFOCT stacks in a completely artifact-free manner for cornea \cite{Mazlin_18} and retina \cite{Xiao_18,Mece_spie_19}. We also proposed a method based on the cumulative sum to enhance non-stationarities in temporal signals which led to an SNR factor increase of 1.9 on average for ex vivo samples and 3 on our in vivo data set. These general techniques could be applied to any other imaging modality with sub-diffraction phase sensitivity.

\section*{Funding}
HELMHOLTZ grant, European Research Council (ERC) (610110).

\section*{Acknowledgments}
The author would like to thank LLTech SAS for sharing its raw data, especially Émilie Benoit and Louis Dutheil for carrying out in vivo and clinical experiments. The author is also grateful to Olivier Thouvenin, Pedro Mecê, Kassandra Groux, Viacheslav Mazlin, Mathias Fink, Claude Boccara and Kate Grieve for fruitful discussions and valuable comments regarding this paper. The data and algorithms used during
the current study are available from the corresponding author upon reasonable request.

\bibliographystyle{unsrt}  
\bibliography{references}  

\end{document}